%% file: Dragone.TEX
\documentclass{article}
\input{prepicte}
\input{pictex}
\input{postpict}

\begin{document}
\title{The Special Relativistic Equivalence Principle: \\ Gravity Theory's Foundation}        
\author{Kenneth Nordtvedt\\Northwest Analysis\\118 Sourdough Ridge, Bozeman MT 59715 USA\\ \it{kennordtvedt@imt.net}}      
\maketitle

\begin{center}
\section*{Abstract}
\end{center}
\begin{quote}
Einstein's Equivalence Principle asserts that physical phenomena occuring in a laboratory which undergoes constant acceleration through gravity-free inertial space should be identical in all respects to that which occurs in local gravity. By  incorporating special relativity theory into an extended version of the Equivalence Principle --- the Special Relativistic Equivalence Principle (SREP) --- {\it post-Newtonian} gravitational phenomena in addition to that originally predicted by Einstein are required (predicted) such as {\it geodetic} precession of local inertial frames which move non-radially in gravitational free fall, precession of Mercury's perihelion, and {\it gravitomagnetic} precession of inertial frames and forces between pairs of mutually moving masses. This poses the historical question --- why were not these phenomena predicted by Einstein in the years 1907-1911? In addition to the predicted precessions, the unique $1/c^2$ order dynamical equations for clock rates and motion of both bodies and light in {\it local} gravity, are derived which guarantee fulfillment of the SREP.    
\end{quote}

\section{Introduction}       

When Einstein formulated his grand hypothesis, the Equivalence Principle (EP), and then used that principle to make his two classic predictions --- that gravity deflects light and alters clock rates --- his arguments rested on only the most rudimentary feature of his special relativity theory; he essentially employed Newtonian physics. A light ray (illustrated in figure 1 by the finely dotted line) leaves an upwardly accelerating floor at initial angle $+\phi$, and it again meets the floor at a later time $T$ and at horizontal distance $L$ as determined from the two Newtonian equations
\begin{figure}
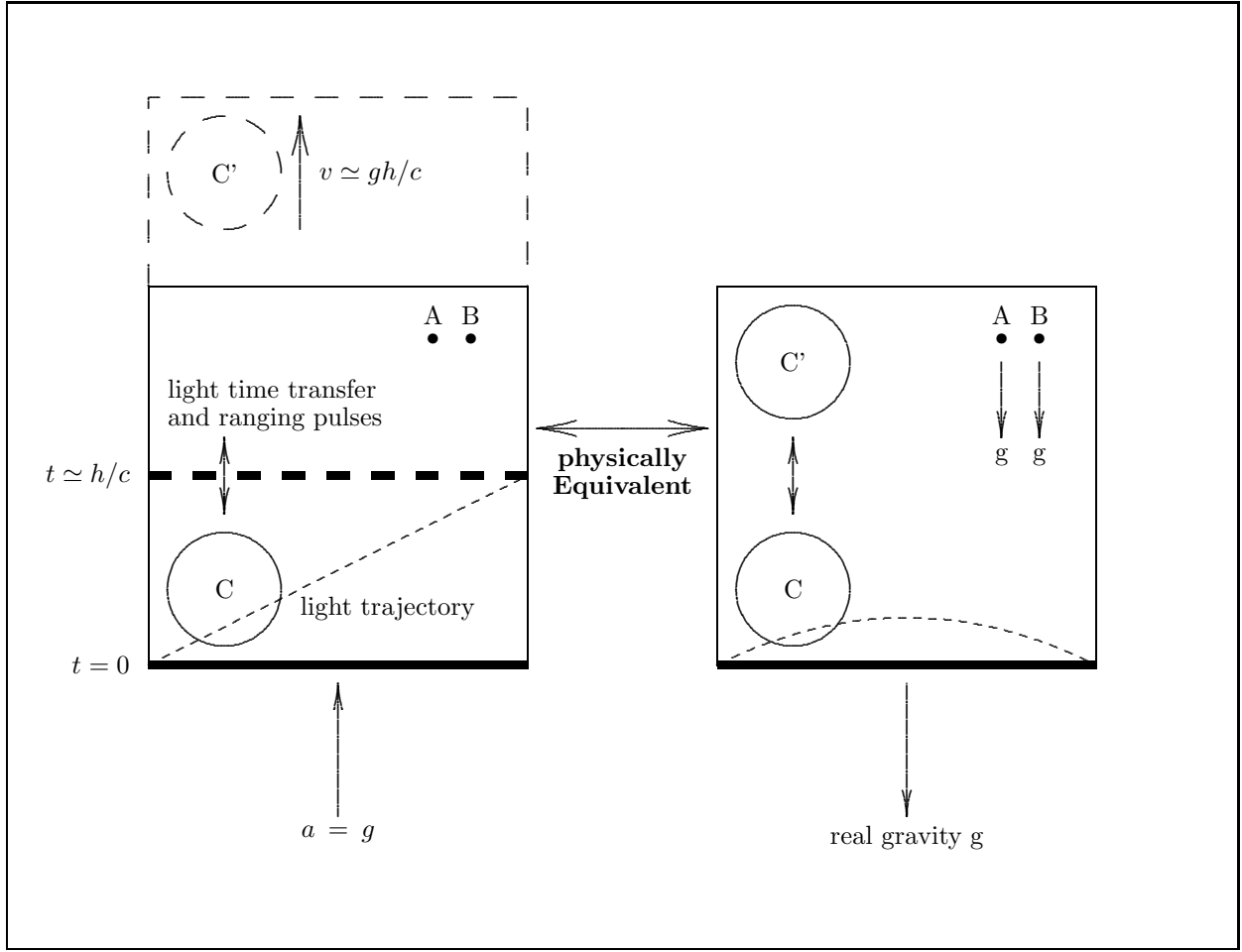

\beginpicture
\setcoordinatesystem units <1in,1in>
\setplotarea x from -3.25 to 3.25, y from -1 to 4
\plotheading{\Large \bf Einstein's Original Equivalence Principle Arguments}
\putrectangle corners at -3.25 -1 and 3.25 4
\putrectangle corners at -2.5 .5 and -.5 2.5
\putrectangle corners at 2.5 .5 and .5 2.5
\linethickness=3pt
\putrule from -2.5 .5 to -.5 .5
\putrule from .5 .5 to 2.5 .5
\setdashes <8.5pt>
\putrule from -2.5 1.5 to -.5 1.5
\setsolid
\linethickness=.4pt
\put {$t=0$} [r] at -2.6 .5
\put {$t\simeq h/c$} [r] at -2.6 1.5
\arrow <10pt> [.1,.33] from -1.5 -.3 to -1.5 .4
\put {$a\:=\:g$} [t] at -1.5 -.35
\arrow <10pt> [.1,.33] from 1.5 .4 to 1.5 -.3
\put {real gravity g} [t] at 1.5 -.35
\circulararc 360 degrees from -2.4 .9 center at -2.1 .9
\circulararc 360 degrees from .6 .9 center at .9 .9
\circulararc 360 degrees from .6 2.1 center at .9 2.1
\put {C} at -2.1 .9
\put {C} at .9 .9
\put {C'} at .9 2.1
\arrow <10pt> [.1,.33] from -2.1 1.3 to -2.1 1.7
\arrow <10pt> [.1,.33] from .9 1.3 to .9 1.7
\arrow <10pt> [.1,.33] from -2.1 1.4 to -2.1 1.3
\arrow <10pt> [.1,.33] from .9 1.4 to .9 1.3
\put {A} [b] at -1 2.3
\put {B} [b] at -.8 2.3
\put {A} [b] at 2 2.3
\put {B} [b] at 2.2 2.3
\put {$\bullet$} [t] at -1 2.25
\put {$\bullet$} [t] at -.8 2.25
\put {$\bullet$} [t] at 2 2.25
\put {$\bullet$} [t] at 2.2 2.25
\put {g} at 2 1.6
\put {g} at 2.2 1.6
\arrow <10pt> [.1,.33] from 2 2.1 to 2 1.7
\arrow <10pt> [.1,.33] from 2.2 2.1 to 2.2 1.7
\arrow <15pt> [.15,.4] from 0 1.75 to -.45 1.75
\arrow <15pt> [.15,.4] from 0 1.75 to .45 1.75
\put {$\bf physically$} [t] at 0 1.65
\put {$\bf Equivalent$} [t] at 0 1.50
\put {light time transfer} [l] at -2.4 1.95
\put {and ranging pulses} [l] at -2.4 1.8
\put {C'} at -2.1 3.1
\arrow <15pt> [.15,.4] from -1.7 2.8 to -1.7 3.4
\put {$v\simeq gh/c$} [l] at -1.6 3.1
\put {light trajectory} [l] at -1.7 .8
\setdashes <8.5pt>
\circulararc 360 degrees from -2.4 3.1 center at -2.1 3.1
\setlinear \plot -2.5 2.5  -2.5 3.5  -.5 3.5  -.5 2.5 /
\setdashes <3pt>
\setlinear \plot -2.5 .5 -.5 1.5 /
\setquadratic \plot .5 .5  1.5 .75 2.5 .5 /
\endpicture
\caption{In the accelerating left box; 1) the floor reaches bodies $A$ and $B$, both at rest in inertial space, at the same time, 2) the floor's right edge has accelerated upwards to meet the light ray, and 3) light pulses sent out by each tick of clock $C$ (anchored to the floor of the box) are received at a slower rate by clock $C'$ (anchored to the ceiling of the box) because of the latter clock's upward motion acquired during the light pulses' times of flight; and the light pulses can be reflected or transponded back to clock $C$. If {\it equivalent} phenomena is to occur in the right box which is at rest in gravity, then 1) bodies $A$ and $B$ must fall at precisely identical rates, 2) light is deflected by gravity, 3) clock $C$ ticks slower than clock $C'$ by virtue of its different location in a gravitational potential, and 4) the round trip ranging time measured by clock $C$ is less than $2h/c$.}
\end{figure}
\[
c\,T\;\sin \phi = \frac{1}{2} g\, T^2\hspace{.3in}and\hspace{.3in}
c\,T\;\cos \phi = L
\]
On reunion at time $T$ the light ray makes a descending angle $-\phi$ with respect to the floor; the rate per unit time for the deflection of that light ray with respect to the floor is then (in the small $\phi$ limit) $d\phi/dt\cong g/c$, or expressed as deflection rate per distance traveled, $d\phi/dx\cong g/c^2$.  Light ray pulses are also indicated in Figure 1, propagating between a clock $C$ anchored to the accelerating floor and another clock $C'$ anchored at height $h$ above the floor. The {\it time transfer} relationship between the times the light leaves the former ($t_1$) and arrives at the latter ($t_2$) is obtained from the Newtonian equation
\begin{equation}
\frac{1}{2} g\, t_1^2\,+\,c\, (t_2-t_1)=h\,+\,\frac{1}{2} g\, t_2^2
\end{equation}
which in first approximation yields a relative rate for these times
\[
\frac{d t_2}{d t_1}\simeq 1+\frac{gh}{c^2}
\]
If round trip ranging experiments using light had been contemplated by Einstein a century ago, he could also have predicted the local outcome of such ranging measurements by adding to equation (1) a relationship for the light's return trip
\begin{equation}
h\,+\,\frac{1}{2}g\,t_2^2-c\,(t_3-t_2)=\frac{1}{2}g\, t_3^2
\end{equation}
which when added to the outbound time gives the round trip's total elapsed time
\[
t_3-t_1\cong \frac{2h}{c}-\frac{gh^2}{c^3}
\]
This EP-derived ranging time for {\it local} experiments is substantiated in metric theories of gravity such as General Relativity and its scalar-tensor variations. The EP predictions of light's deflection in gravity have been claimed by some to be no more than earlier predictions of mechanistic deflection of light corpuscles traveling at the finite speed $c$.  This mechanistic viewpoint, however, would predict a speeding up of light as it approached matter, not the slowing obtained from the EP \cite{fn5}.

The third phenomena illustrated in figure 1 consists of the generally different bodies $A$ and $B$ which are at rest and located side by side in inertial space.  The upwardly accelerating floor then meets both of these bodies simultaneously; indeed it was Einstein's contemplation of this identity of free fall which led him to his principle.

Requiring these observational results to also occur in gravity by virtue of the EP, the interpretations must now be that the local gravitational acceleration $g$ 1) deflects a transversely propagating light ray,  2) changes clock frequencies $f$ with altitude $h$, and 3) increases the speed of light (as measured by a ground clock) by the  previously derived rates,
\[
\frac{d \phi}{d x}= \frac{1}{f} \frac{df}{dh}= \frac{1}{c}\frac{dc}{dh}=\frac{g}{c^2}
\] 
and 4) different bodies $A$ and $B$ {\it fall} in gravity at precisely identical rates. Special relativity played almost no role in arriving at these conclusions.  

But the EP can predict a number of additional novel phenomena.  By fully utilizing special relativity when exploring implications of the EP, converting it into the special relativistic Equivalence Principle (SREP), further effects can be predicted which include1) {\it geodetic} precession of a body's inertial orientation as it free-falls non-vertically in gravity, 2) a relativistic ($1/c^2$ order) contribution to precession of the major axes of gravitational orbits (such as Mercury's), and 3) a  {\it gravitomagnetic} precession of a body's inertial orientation by virtue of a moving source of gravity, as well as a general gravitational interaction between mutually moving masses and between moving mass and light. 

The derivation of these new consequences of {\it equivalence} follows the spirit of the original EP arguments. Novel phenomena are first derived as they occur in gravity-free, accelerated laboratories. To analyze body and light ray trajectories, clock rates, and behavior of other experimental devices, we set up a {\it master} inertial frame with its observer and clock at rest, and from that perspective the calculations of clock, body, and light behaviors can be performed.  In this gravity-free inertial frame light rays travel along  straight lines at unique speed $c$, free bodies move at constant velocities, and arbitrarily moving clocks 'tick' at the special relativistic {\it proper} rate $d\tau$
\begin{equation}
d\tau\;=\;dt\;\sqrt{1-v(t)^2/c^2}
\end{equation}
expressed in terms of the rate $dt$ of a clock at rest in the master inertial frame. A 'ground' floor of clocks are synchronously given equal and constant (properly measured by accompanying accelerometers) upward accelerations.  To keep the interpretations of various measurable phenomena as straightforward and free of controversy as possible, the experimental observables are confined to measurements made on this ground floor of accelerating clocks (later, of course, an equivalent array of clocks is deployed on the actual {\it ground} in a local gravitational field).  Special relativity's Lorentz transformation, used to relate event coordinates as measured in two inertial frames which move at constant velocity  relative to each other, is needed; in the case of a transformation to a frame moving at speed $v$ in the $y-direction$, for example, new coordinates are related to original ones by
\begin{eqnarray}
dt'\;&=&\;\gamma\;\left(dt-v\:dy/c^2\right) \nonumber \\
dy'\;&=&\;\gamma\;\left(dy-v\:dt\right) \nonumber \\
dx'\;&=&\;dx \nonumber \\
dz'\;&=&\;dz\hspace{1in}with\;\gamma\;=\frac{1}{\sqrt{1-v^2/c^2}} \nonumber
\end{eqnarray} 

The types of {\it gedanken} experiments analyzed in this gravity-free situation of a ground floor of upwardly accelerating clocks are shown in the bottom picture of Figure 4.  Both  bodies and light rays are  considered which are given free trajectories initially leaving the accelerating ground floor.  The bodies may carry clocks and have extension (orientation). At future times there will be reunions of the body (clock) trajectories and light trajectories with that of the upwardly accelerating ground floor. Various measurable quantities are then recorded at these reunion events; such measurements include the elapsed proper times of various clocks, the body orientations, horizontal locations of reunions, etc. {\bf The SREP then requires identical results for the same measurable quantities in gravity, as shown in the top picture of Figure 4. In order to achieve this identity of results, unique gravity-induced modifications to the speed of light function, to the body equation of motion, and to the clock rate function are determined, and rotations of an inertial rod with respect to the ground during free fall motion are required.}   
\begin{figure}
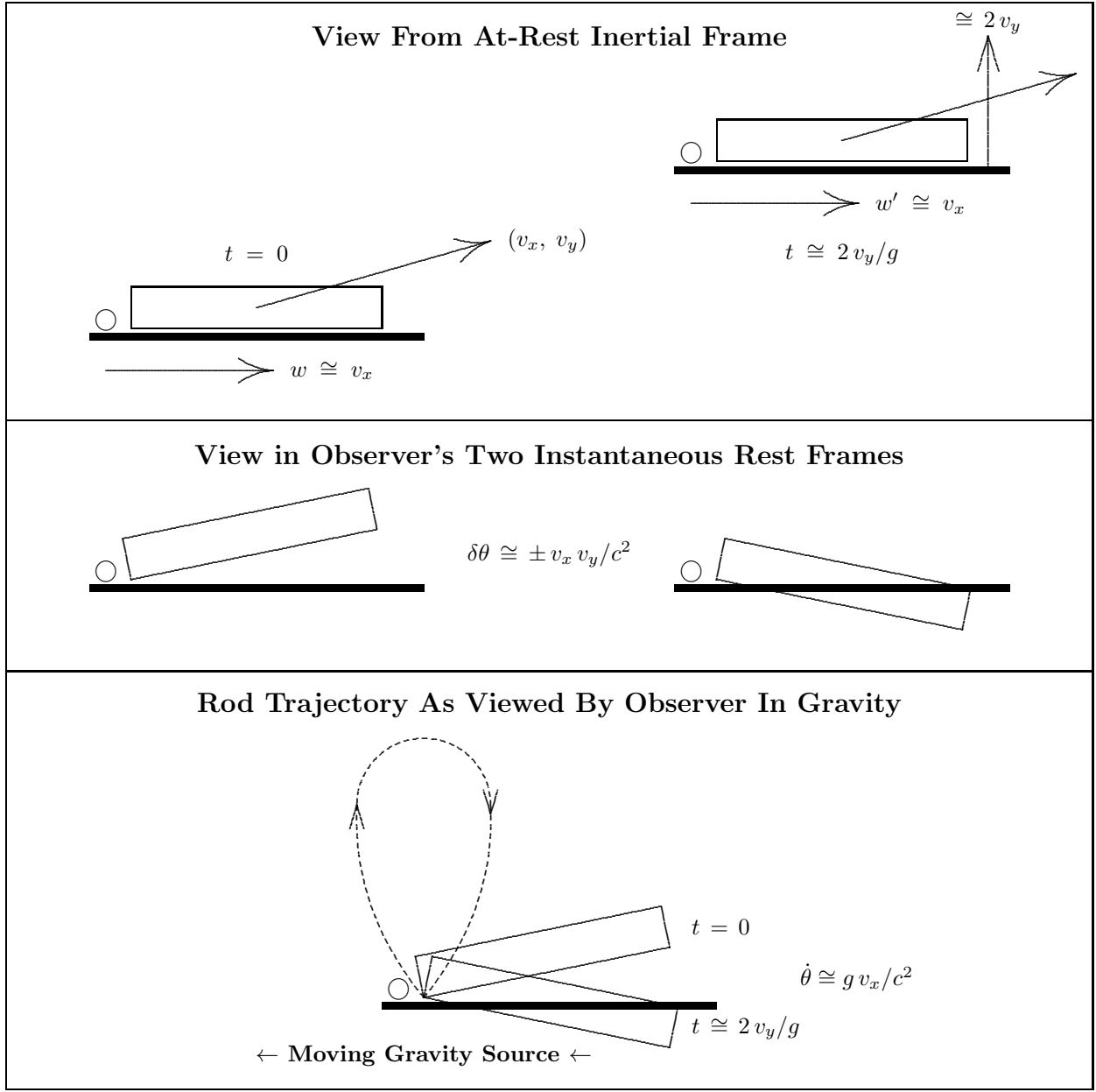

\beginpicture
\setcoordinatesystem units <1in,1in>
\setplotarea x from -3.25 to 3.25, y from 0 to 6.5
\plotheading{\Large \bf Gravitomagnetic Precession From The Equivalence Principle}
\putrectangle corners at -3.25 0 and 3.25 6.5
\putrule from -3.25 2.5 to 3.25 2.5
\putrule from -3.25 4 to 3.25 4
\putrectangle corners at -2.5 4.55 and -1 4.8
\putrectangle corners at 1 5.55 and 2.5 5.8
\startrotation by .98 .2 about -2.5 3.05
\setlinear \plot -2.5 3.05 -1 3.05 -1 3.3 -2.5 3.3 -2.5 3.05 /
\stoprotation
\startrotation by .98 .2 about -.75 .55
\setlinear \plot -.75 .55 .75 .55 .75 .8 -.75 .8 -.75 .55 /
\stoprotation
\startrotation by .98 -.2 about 1 3.05
\setlinear \plot 1 3.05 2.5 3.05 2.5 3.3 1 3.3 1 3.05 /
\stoprotation
\startrotation by .98 -.2 about -.75 .55
\setlinear \plot -.75 .55 .75 .55 .75 .8 -.75 .8 -.75 .55 /
\stoprotation
\put {$t\,\cong\,2\,v_y/g$} [l] at .85 .37
\put {$t\,=\,0$} [l] at .85 .97
\put {$\dot{\theta}\cong g\,v_x/c^2$} [l] at 1.5 .67
\setdashes <2pt>
\circulararc 180 degrees from -.35 1.7 center at -.75 1.7
\setquadratic \plot -.75 .55 -.435 1.1 -.35 1.7 /
\setquadratic \plot -.75 .55 -1.065 1.1 -1.15 1.7 /
\setsolid
\arrow <10pt> [.3,.6] from -.35 1.66 to -.35 1.65
\arrow <10pt> [.3,.6] from -1.15 1.69 to -1.15 1.7
\put {$\bigcirc$} at -2.65 3.1
\put {$\bigcirc$} at .85 3.1
\put {$\bigcirc$} at -2.65 4.6
\put {$\bigcirc$} at .85 5.6
\put {$\bigcirc$} at -.9 .6
\put {$t\;=\;0$} at -1.75 5
\put {$t\;\cong\;2\,v_y/g$} at 1.75 5
\arrow <15pt> [.2,.7] from  1.75 5.675 to 3.15 6.075
\arrow <15pt> [.2,.7] from -1.75 4.675 to -.35 5.075
\arrow <15pt> [.2,.7] from -2.65 4.3 to -1.65 4.3
\arrow <15pt> [.2,.7] from .85 5.3 to 1.85 5.3
\put {$(v_x,\;v_y)$} [l] at -.25 5.075
\put {$\delta\theta\,\cong\,\pm\,v_x\,v_y/c^2$} at 0 3.2
\put {$w\;\cong\;v_x$} [l] at -1.55 4.3
\put {$w'\;\cong\;v_x$} [l] at 1.95 5.3
\put {\large \bf View in Observer's Two Instantaneous Rest Frames} at 0 3.8
\put {\large \bf View From At-Rest Inertial Frame} at 0 6.3
\put {\large \bf Rod Trajectory As Viewed By Observer In Gravity} at 0 2.3
\arrow <15pt> [.3,.6] from 2.625 5.5 to 2.625 6.3
\put {$\cong\,2\,v_y$} at 2.625 6.38
\put {\bf $\leftarrow$ Moving Gravity Source $\leftarrow$} at -.75 .2
\linethickness=3pt
\putrule from -2.75 3 to -.75 3
\putrule from .75 3 to 2.75 3
\putrule from -2.75 4.5 to -.75 4.5
\putrule from .75 5.5 to 2.75 5.5
\putrule from -1 .5 to 1 .5
\endpicture
\caption{The top scene shows an upwardly accelerating floor and a non-rotating rod moving freely through gravity-free space.  Floor and rod meet twice, and an observer moves at constant proper velocity along the floor to be present at both events.  The middle scene shows the meetings in the two instantaneous rest frames of the observer. The relativity of simultaneity in the Lorentz-transformation for time results in different rotations of the rod in the two events. The Equivalence Principle calls for the same observable outcomes in gravity; this predicts gravitomagnetic rotation relative to the ground of an {\it inertially non-rotating} rod due to a moving source of the gravity.  The bottom scene also shows that the rod's gravitational free fall trajectory is not vertical as viewed from the observer.  This specifies the local gravitomagnetic contribution to the gravitational equation of motion.} 
\end{figure}
    
\begin{figure}
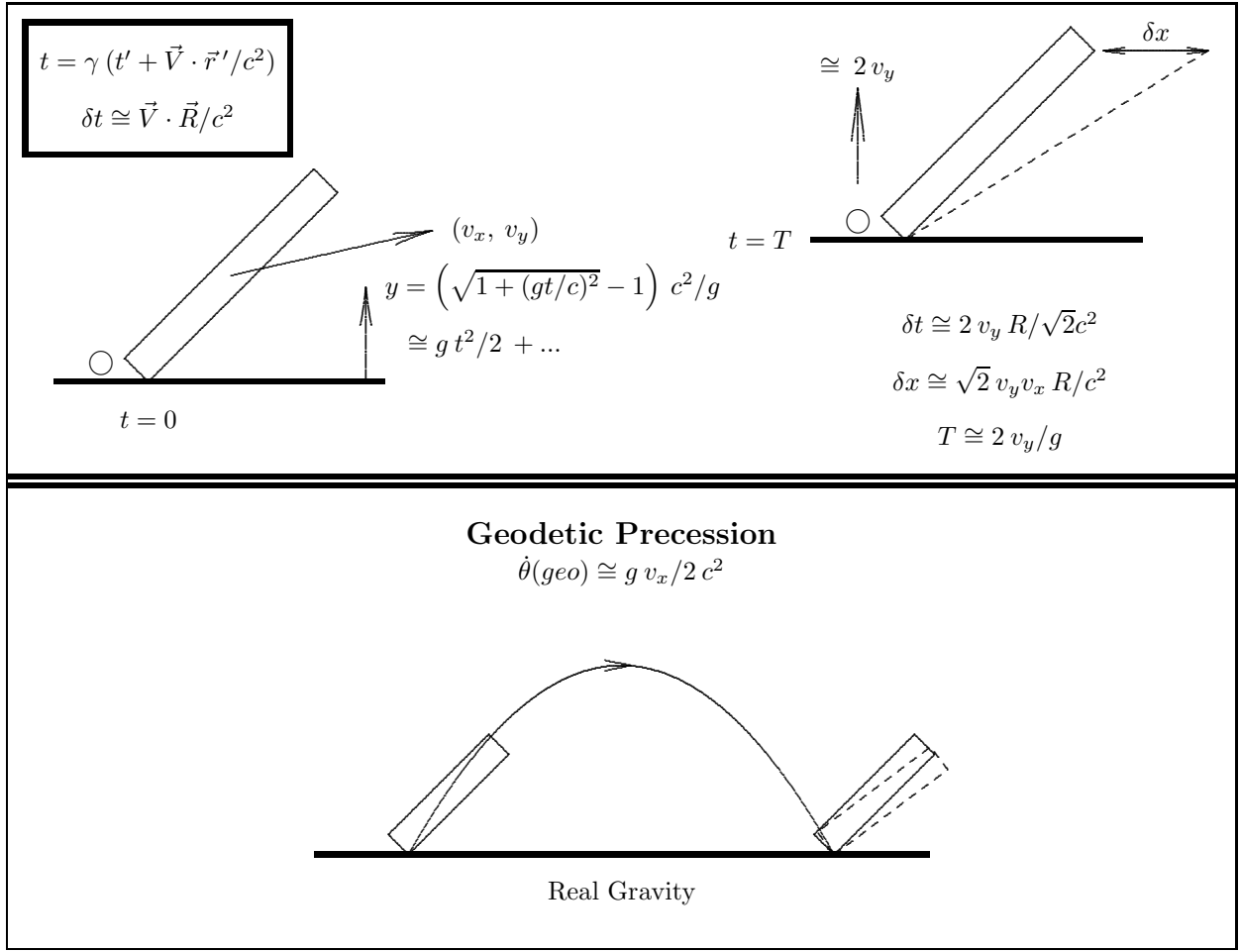

\beginpicture
\setcoordinatesystem units <1in,1in>
\setplotarea x from -3.25 to 3.25, y from -3 to 2
\putrectangle corners at -3.25 -3 and 3.25 2
\plotheading{\Large \bf Geodetic Precession From The Equivalence Principle}
\put {$\delta t\cong 2\: v_y\,R /\sqrt{2}c^2$} at 2 .3
\put {$\delta x\cong \sqrt{2}\: v_y v_x \,R /c^2$} at 2 0
\put {$T\cong 2\:v_y/g$} at 2 -.3
\arrow <15pt> [.12,.3] from -1.35 0 to -1.35 .5
\arrow <15pt> [.12,.3] from -2.0625 .5625 to -1 .8
\put {$(v_x,\:v_y)$} [l] at -.9 .8
\put {$y=\left(\sqrt{1+(gt/c)^2}-1\right)\:c^2/g$} [l] at -1.25 .5
\put {$\cong g\:t^2/2\:+...$} [l] at -1.13 .2
\put {$t=\gamma\: (t'+\vec{V}\cdot\vec{r}\:'/c^2)$} at -2.45 1.7
\put {$\delta t\cong\vec{V}\cdot\vec{R}/c^2$} at -2.45 1.4
\linethickness=2pt
\putrule from -3.25 -.5 to 3.25 -.5
\putrule from -3.25 -.55 to 3.25 -.55
\putrectangle corners at -3.15 1.9 and -1.75 1.2
\putrule from -3 0 to -1.25 0
\putrule from 1 .75 to 2.75 .75
\putrule from -1.625 -2.5 to 1.625 -2.5
\put {$t=T$} [r] at .9 .75
\linethickness=.4pt
\setquadratic \plot -1.125 -2.5 0 -1.5 1.125 -2.5 /
\arrow <10pt> [.2,.4] from 0 -1.5 to .05 -1.5
\put {\large \bf Geodetic Precession} at 0 -.8
\put {$\dot{\theta}(geo)\cong g\:v_x/2\:c^2$} at 0 -1
\setlinear \plot -2.5 0 -1.5 1 -1.625 1.125 -2.625 .125 -2.5 0 /
\startrotation by .707 .707 about -1.125 -2.5
\setlinear \plot -1.125 -2.5 -.375 -2.5 -.375 -2.35 -1.125 -2.35 -1.125 -2.5 /
\stoprotation
\startrotation by .707 .707 about 1.125 -2.5
\setlinear \plot 1.125 -2.5 1.875 -2.5 1.875 -2.35 1.125 -2.35 1.125 -2.5 /
\stoprotation
\setdashes <3pt>
\startrotation by .8 .6 about 1.125 -2.5
\setlinear \plot 1.125 -2.5 1.875 -2.5 1.875 -2.35 1.125 -2.35 1.125 -2.5 /
\stoprotation
\setsolid
\put {Real Gravity} at 0 -2.7
\put {$t=0$} at -2.5 -.2

\put {$\bigcirc$} at -2.75 .1
\setlinear \plot 1.5 .75 2.5 1.75 2.375 1.875 1.375 .875 1.5 .75 /
\put {$\bigcirc$} at 1.25 .85
\arrow <20pt> [.12,.24] from 1.25 1.05 to 1.25 1.55
\put {$\cong\: 2\,v_y$} at 1.25 1.65
\arrow <8pt> [.15,.3] from 2.55 1.75 to 3.1 1.75
\arrow <8pt> [.15,.3] from 3.1 1.75 to 2.55 1.75
\put {$\delta x$} at 2.825 1.85
\setdashes <3pt>
\setlinear \plot 3.1 1.75 1.5 .75 /
\setsolid
\endpicture
\caption{The top scenes shown from perspective of a master inertial frame show an intrinsically non-rotating rod both when it leaves and when it again meets an upwardly accelerating floor.  After Lorentz transformations to the instantaneous rest frames of the observer (fixed on the floor) indicated by symbols $\bigcirc$, the orientations of the rod relative to the floor are shown by the dotted rod.  The key {\it time} Lorentz transformation responsible for the rod reorientation is shown in upper left corner of the figure.  Assuming the SREP, the figure's bottom scene shows the same rotation of the rod in gravity, but which now must be interpreted as  {\it geodetic} rotation of the inertial frame which moves through gravity with the rod.}
\end{figure}

Consider a rod which travels at constant velocity and without rotation through gravity-free inertial space. (``Non-rotation'' of the rod can be established, for instance, by attached accelerometers which record no centrifugal forces.)  As shown in Figure 4, the trajectory of this rod  is twice crossed by that of an upwardly accelerating ground floor of the non-inertial laboratory. In the instantaneous rest frames of those two crossing events the orientations of the rod with respect to the ground are determined and found to differ. When the SREP is then invoked and a rod free falling in gravity (and free of absolute rotation) is considered, this same change of orientation will be required, but that rotation must now be interpreted as a precession of the rod's inertial orientation by virtue of its motion through the local gravity --- {\it geodetic precession},  or by virtue of the motion of the source of gravity --- {\it gravitomagnetic precession}.

\begin{figure}
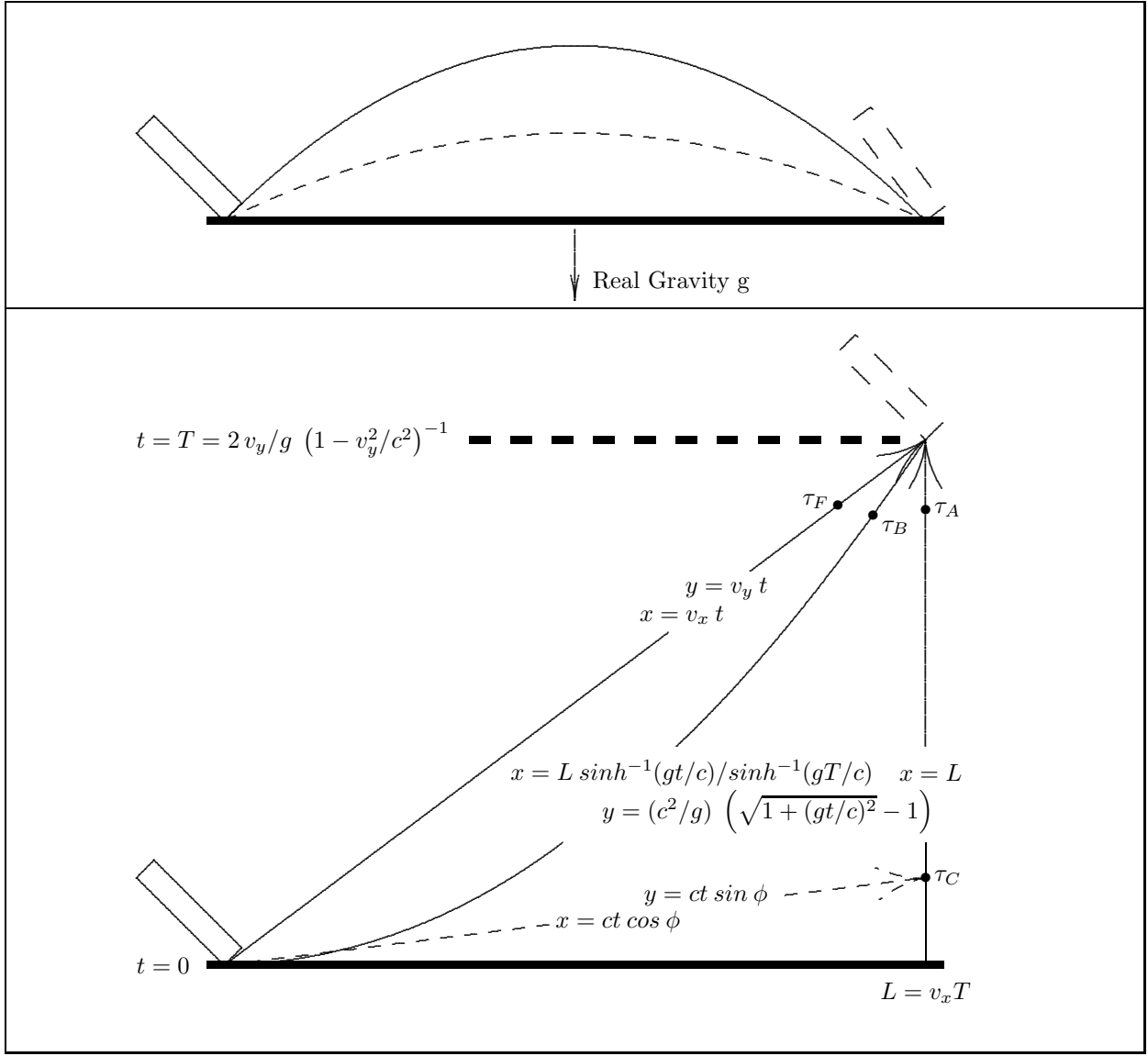

\beginpicture
\setcoordinatesystem units <1in,1in>
\setplotarea x from -3.25 to 3.25, y from -.5 to 5.5
\putrectangle corners at -3.25 -.5 and 3.25 5.5
\plotheading{\large \bf Clock/Rod and Light Ray Leave and Rejoin the 'Ground': Two Views}
\setlinear \plot -2 0  -1.9 .1  -2.4 .6 -2.5 .5 -2 0 /
\setlinear \plot -2 4.25 -1.9 4.35 -2.4 4.85 -2.5 4.75 -2 4.25 /
\linethickness=3pt
\putrule from -2.1 0 to 2.1 0
\putrule from -2.1 4.25 to 2.1 4.25
\setdashes <8.5pt>
\putrule from -.6 3 to 1.85 3
\setsolid
\linethickness=.4pt
\putrule from -3.25 3.75 to 3.25 3.75
\setquadratic \plot -2 4.25 0 5.25 2 4.25 /
\setquadratic \plot -2 0 -1 .1875 -.07 .7 /
\setquadratic \plot .58 1.25 1.266 2 2 3 /
\put {$\bullet$} at 1.7 2.567
\put {$\bullet$} at 2 2.6
\put {$\bullet$} at 1.5 2.625
\put {$\bullet$} at 2 .5
\put {\bf $\tau_C$} [l] at 2.05 .5
\put {\bf $\tau_A$} [l] at 2.05 2.6
\put {\bf $\tau_B$} [l] at 1.75 2.5
\put {\bf $\tau_F$} [r] at 1.45 2.65
\put {$t=0$} [l] at -2.5 0
\put {$t=T=2\,v_y/g\;\left(1-v_y^2/c^2\right)^{-1}$} [l] at -2.5 3
\put {$x=L\:sinh^{-1}(gt/c)/sinh^{-1}(gT/c)$} [r] at 1.7 1.1
\put {$x=L$} [l] at 1.85 1.1
\arrow <20pt> [.2,.67] from 1 2.25 to 2 3
\setlinear \plot -2 0 .533 1.9 /
\arrow <20pt> [.2,.67] from 2 1.25 to 2 3
\arrow <10pt> [.1,.33] from 0 4.2 to 0 3.8
\put {Real Gravity g} [l] at .1 3.9
\putrule from 2 0 to 2 .7
\put {$x=v_x\:t$} [r] at .85 2
\put {$y=v_y\:t$} [r] at 1.1 2.15
\put {$x=ct\:cos\:\phi$} [r] at .6 .25
\put {$y=ct\:sin\:\phi$} [r] at 1.1 .4
\put {$L=v_x T$} [t] at 2 -.1
\put {$y=(c^2/g)\;\left(\sqrt{1+(gt/c)^2}-1 \right)$} [t] at 1.1 1
\setdashes
\arrow <20pt> [.2,.67] from 1.2 .4  to 2 .5
\setlinear \plot -2 0 -.1 .238 /
\setdashes <10pt>
\setlinear \plot 2 3  2.1 3.1  1.6 3.6  1.5 3.5  2 3 /
\setlinear \plot 2 4.25 2.113 4.335 1.688 4.9 1.575 4.815 2 4.25 /
\setdashes
\setquadratic \plot -2 4.25 0 4.75 2 4.25 /
\endpicture
\caption{The bottom view of events is as seen by a master inertial observer at rest in gravity-free space.  At time $t\,=\,0$ a light ray (dashed line)is launched at angle $\phi$, and a non-accelerating, non-rotating rod with clock is launched at angle $tan^{-1}(v_y/v_x)$ (solid line), and then ray and rod meet an upwardly accelerating 'ground' clock at the latter's times {\bf $\tau_C$} and {\bf $\tau_A$}, respectively.  Another 'ground' clock moves at constant proper speed $w$ to the right to also meet the rod/clock at the reunion event. The non-accelerating (free falling) clock records the time {\bf $\tau_F$} for its reunion event, and the right-moving 'ground' clock records the time {\bf $\tau_B=\tau_A\:\sqrt{1-w^2/c^2}$}. The trajectories of the three clocks and the light ray as recorded in the master inertial frame are indicated. The top view shows the same physical events occuring in gravity. The SREP requires all observables such as the clock readings at the reunions, etc., to have identical values in the two situations.}
\end{figure}

\section{Gravitomagnetic Precession Due to Moving Gravity Source}

As viewed from a master inertial frame (top panel of Figure 2), at time $t\,=\,0$ a horizontal rod leaves a floor with horizontal velocity component $v_x$ and vertical velocity $v_y$ \cite{fn1}.  An observer travels along the floor at constant horizontal {\it proper} speed $w$ selected so as to arrive at the future reuniting event of rod and floor.  The floor accelerates upward as $y\,\cong\,g\,t^2/2$.  In the $t'\,=\,0$ rest frame of this observer, the {\it time} Lorentz transformation indicates different times as measured in the master inertial frame for the two ends of the rod
\begin{equation}
t\;=\;\gamma\:\left(t'\,+\,\vec{V}\cdot\vec{r}\,'/c^2\right)\quad with\;\gamma\;=\;\frac{1}{\sqrt{1-v^2/c^2}}
\end{equation}
with the right side of the rod having the later time $t$ value.  With the rod initially moving up from the floor, the middle panel of Figure 2 indicates the rod's initial orientation as seen in the observer's rest frame.  The difference between rod orientation angles  seen in the two frames is readily evaluated to be $v_xv_y/c^2$ in leading relativistic order.

The upwardly accelerating floor meets the rod again at time $T\,\cong\,2\,v_y/g$ by which time the floor is traveling upward at speed of about $2\,v_y$.  In the instantaneous rest frame of the observer at this second meeting of floor with rod, the Lorentz transformation given in equation (4) can again be used to find that the master inertial frame time for the right end of the rod is later than that for the rod's left end.  But since the rod is now traveling down relative to the floor at  speed of about $v_y$, the relationship between rod orientations as seen in the master inertial frame and instantaneous rest frame of the observer is now reversed as also shown in the middle panel of Figure 2; the latter orientation is now turned down from the horizontal orientation by angle which is again $v_xv_y/c^2$.  Dividing by the total elapsed time $T$ between the two events, one obtains a precession rate for the rod relative to the floor
\begin{equation}
\dot{\Theta}_{LT}\;\cong\;\frac{g\,v_x}{c^2}
\end{equation}
labeled ``$LT$'' is recognitionof the pioneering work of Lense and Thirring concerning this precession in General Relativity theory \cite{LT}.
As seen from a frame of reference at rest with the observer, a rod is launched (almost) vertically into gravitational free fall.  Upon return to the ground, the rod has rotated while nevertheless not experiencing internal centrifugal accelerations.  An observer moving at constant velocity along an upwardly accelerating floor detects his motion;  there is a preferred frame on this floor established by special relativity.  In gravity, on the other hand, the only available explanation for this rotation is the observer's presence in a gravitational field and the leftward horizontal motion of the gravitational source relative to the observer's frame. {\bf In proximity to a moving source of gravity, the local inertial frame must rotate! } The slight non-verticality of the free fall trajectory which is another consequence of gravitomagnetism is discussed in Section 4. 

\section{Geodetic Precession Due to Motion Through Gravity}

The top panel of Figure 3 illustrates the geodetic precession case.  Two observers are fixed on the floor; one is located where a rod is launched upward from the floor and another is located where the rod again meets the upwardly accelerating floor.  It is convenient to orient the rod at $45\;degrees$ with respect to the floor; at this orientation the two different Lorentz contractions of the rod seen in the instantaneious rest frames of the two observers produce identical angular change of the rod with the floor, and the discussion is simplified.  The instantaneous rest frame of the observer at the $t\,=\,0$ event coincides with the master inertial frame, so the solid rod indicates the orientation measured in that observer's instantaneous rest frame.  But the second observer is moving upward at speed of about $2\,v_y$ when the second meeting of rod and floor occurs.  Therefore the Lorentz transformation of times given in equation (4) must again be used to understand this latter event. At some time in the second observer's instantaneous rest frame for the meeting, the {\it time} Lorentz transformation measures a time difference for the rod's two ends as seen in the master inertial frame of $\delta t\,\cong\,2\,v_y\,R/\sqrt{2}c^2$ with $R$ being the length of the rod. In this time interval the right end of the rod moves distance $\delta x\,\cong\,\sqrt{2}\,v_x v_y\,R/c^2$ further to the right, thereby decreasing the angle between the rod and the floor in amount $v_x\,v_y/c^2$.  Dividing by the total time $T$ between these events then yields the precession rate relative to the floor
\begin{equation}
\dot{\Theta}_{geodetic}\;\cong\;\frac{1}{2}\frac{g\,v_x}{c^2}
\end{equation}
Since the observers in this case are at rest with respect to the source of gravity, this precession of the inertial rod must be explained as due to the motion of that rod transversely through the gravitational field, i.e., {\it geodetic} precession.

\section{General Consideration Of The Observables}

A rod with clock moves at constant velocity and without rotation through the master inertial frame as shown in Figure 4. At $t=0$ its lower end "1" leaves the floor (ground) which is upwardly accelerating. Expressed in the master inertial frame which for convenience is selected to coincide with the instantaneous rest frame of the floor at $t=0$, the trajectories of the rod's two ends are
\begin{eqnarray}
x_1(t) = v_x t\; \mbox { and }\;x_2(t)=x_1(t)+X \\
y_1(t) = v_y t\; \mbox { and }\;y_2(t)=y_1(t)+Y
\end{eqnarray}
with $X,\; Y,\; v_x$ and $v_y$ all positive \cite{fn4}.  The `fixed ground' clocks have no horizontal motion and the common vertical motion
\begin{equation}
y(t)=\frac{c^2}{g}\; \left(\sqrt{1+(gt/c)^2}-1 \right)
\end{equation}
which manifests constant acceleration $g$ as measured by accelerometers accompanying these clocks. The y-motion given in equation (9) catches up with $y_1(t)$ from equation (8) at master inertial frame time
\begin{equation}
T = \frac{2 v_y}{g}\; \frac{1}{1-v_y^2/c^2}
\end{equation}
which event occurs at horizontal location
\begin{equation}
L=v_x\;T
\end{equation}
with the floor moving upward at speed 
\begin{equation}
V = \frac{2 v_y}{1+v_y^2/c^2}
\end{equation}
as measured in the master inertial frame.  The rod's vertical velocity relative to the floor, as measured in the rest frame of the floor at the reunion event, is obtained using the special relativistic transformation rule for velocities:
\[
v_y'=\frac{v_y - V}{1-v_y V/c^2}= - v_y
\]
an unsurprising result. At this reunion event the horizontal velocity of the rod as measured in the instantaneous rest frame of the ground is found equal to its original horizontal velocity, so the trajectory's locally measured arrival angle is the negative of the original locally measured departure angle. 

In the instantaneous rest frame of the floor at reunion with the rod end "1", the master inertial frame event coordinates are
\[
t_1'=\gamma\;T\; \left(1-v_y V/c^2 \right) \hspace{.5in}x_1'=v_x T \hspace{.5in}y_1'=\gamma\; \left(v_y-V \right) T 
\]
with
\[
\gamma=\frac{1}{\sqrt{1-V^2/c^2}}
\]
In this frame, and at the moment its end "1" meets the floor, we also want to know where the rod's other end "2" is at?  From the time transformation of special relativity  we have
\[
t_1'=\gamma\; \left(t_2-V(Y+v_y t_2)/c^2 \right)
\]
which gives
\[
t_2=T+Y \frac{V}{c^2-v_y V}
\]
The location of rod end "2" at that moment is then
\begin{eqnarray}
x_2'&=&X+v_x \left(T+Y \frac{V}{c^2-v_y V} \right) \nonumber \\
y_2'&=&\gamma \left(Y+(v_y-V)\left(T+Y \frac{V}{c^2-v_y V} \right)\right) \nonumber
\end{eqnarray}

The orientations of the rod at the two crossings of the floor can now be compared. Constructing the tangents of the angles the rod makes with the floor in the two instances, measured in each case in the floor's instantaneous rest frame, 
\begin{eqnarray}
\tan \phi = (y_2-y_1)/(x_2-x_1) &=& \frac{Y}{X} \nonumber \\[.4in]
\tan \phi'=(y_2'-y_1')/(x_2'-x_1')&=& \frac{Y}{X}\;\gamma\; 
\frac{c^2-V^2}{c^2+v_x VY/X-v_y V} \nonumber
\end{eqnarray}
the difference between these angles represents change of the rod's orientation relative to floor in a clockwise sense. In the limit of small vertical velocities of the rod, this rotation angle is
\[
\delta\phi \simeq\frac{v_x v_y}{c^2}\;(1-\cos 2\phi)
\]
The $\cos 2\phi$ term of this expression is simply due to the change in the Lorentz contraction of the rod as its velocity components (with respect to the floor) have changed from $(v_x,v_y)$ to $(v_x,-v_y)$. The remaining constant term of the expression is equivalent to a secular precession rate
\begin{equation}
\frac{d\phi}{dt}=\frac{1}{2}\frac{g\, v_x}{c^2}
\end{equation}
which confirms the conclusion in Section 3.  The SREP requires this precession to also occur for an inertial rod which is on a free fall trajectory in gravity. 

How dramatic it would have been in the era 1907-1911 when Einstein had yet no theory of gravity but only his Equivalence Principle, if he had publically predicted not only that inertial frames are local, not global, and undergo free fall acceleration in gravity, but also that if these frames are moving non-radially in that gravity, they must rotate with respect to more distant inertial frames!  It remained until just after Einstein's publication of his complete theory of general relativity for Willem deSitter in 1916 to discover by calculation the full {\it geodetic} precession contribution to the Moon's perigee rotation rate with respect to distant inertial space, one third of which has here been shown to follow from the SREP \cite{wds16}. 

Additional observables can be established by considering a number of clocks, some in free motion, some fixed in position on the upwardly accelerating ground floor, and others  moving at {\it constant proper speed} along the upwardly accelerating ground floor. Each of these clocks undergoes an interval of elapsed proper time which depends on its specific motion in the master inertial frame 
\begin{equation}
d\tau_i=\sqrt{1-v_i(t)^2/c^2}\; dt
\end{equation}
with $dt$ being the elapsed proper time increment of a clock at rest in the master inertial frame. Using the previously derived {\it master} time of reunion of the rod end "1" (also carrying a clock) with the ground, given by equation (10), this free clock on the rod  records this reunion event at an elapsed proper time since launch
\begin{equation}
\tau(v_y,v_x)_F\;=\;T\;\sqrt{1-(v_x^2+v_y^2)/c^2}\;=\;\frac{2v_y}{g}\;
\sqrt{1-(v_x^2+v_y^2)/c^2}\;\frac{1}{1-v_y^2/c^2}
\end{equation}
The trajectory of the fixed ground clocks trajectory is given by equation (9); integrating the proper time expression given by equation (14) then gives that clock's elapsed proper time between the launch event and the reunion event with the free-falling clock 
\begin{equation}
\tau_A\;=\;\int_0^T\sqrt{1-(dy/dt)^2/c^2}\;dt\;=\;
\frac{c}{g}\;\sinh^{-1}\left(gT/c\right)\;=\;
\frac{c}{g}\;\sinh^{-1}\left(\frac{2v_y}{c}\frac{1}{1-v_y^2/c^2}\right)
\end{equation}
which is independent of $v_x$, unlike the case for the elapsed proper time of the free (freely falling) clock. 

A third type of clock permits the interesting variation on this experiment in which the same ground clock records both the launch from and reunion with the ground of the freely falling clock. This is achieved by giving that ground clock an initial velocity $w$ to the right such that it arrives at horizontal location $L$ simultaneously with the freely falling clock. Because of the upward acceleration of the ground, its horizontal velocity does not remain constant as seen in the master inertial frame; it moves according to
\[
dx/dt\;=\;w\;\sqrt{1-(dy/dt)^2/c^2}
\]
which, however, fulfills the requirement that no horizontal force acts on the clock
\[
\frac{d}{dt}\frac{dx/dt}{\sqrt{1-v^2/c^2}}\;=\;0
\]
and that equal intervals of $x$ are traveled per unit of proper time recorded on the horizontally moving clock. Since we want the simultaneous arrival of the {\it free falling} clock and the clock moving along the ground, this requires
\[
\int_0^T (dx/dt)dt\;=\;w\;\int\sqrt{1-(dy/dt)^2/c^2}\;dt\;=\;L
\]
requiring an initial horizontal speed $w$ which is greater than that of the freely falling clock
\[
w\;=\;v_x\;\frac{gT/c}{sinh^{-1}(gT/c)}\;=\;v_x\:\left(1\:+\:\frac{1}{6}\frac{g^2T^2}{c^2}\:+\:...\right)
\]
with $T$ given in equation (10). The proper time of the reunion event as recorded by this moving clock is $\tau(w)_B=\tau_A\;\sqrt{1-w^2/c^2}$.  Since $w$ is in excess of $v_x$, in the frame of reference traveling to the right with this moving clock $B$, the freely falling clock is not launched vertically; it must instead be launched to the left of vertical (see top view in Figure 5) at angle $\Theta\cong -2v_x v_y/3c^2$ (for non-relativistic speeds $v_x$ and $v_y$), and more generally, at an angle 
\[
\tan\Theta\;=\;\frac{v_x-w}{1-wv_x/c^2}\frac{\sqrt{1-w^2/c^2}}{v_y}
\]
These elapsed proper times, $\tau(v_y,v_x)_F$, $\tau_A$, and $\tau(w)_B$, and the horizontal location $L$ of the reunion event from equation (11), are observables which must all be reproduced in the equivalent gravity environment if the SREP is to be fulfilled.

Some of these observables are relevant to the case in which the inertially moving rod is replaced by a light ray. It's trajectory in the master inertial frame is
\[
x=ct\;\cos\phi \hspace{.5in} y=ct\;\sin\phi
\]
The time recorded by a clock at rest in the master inertial frame for the reunion of the light ray with the ground is then
\[
T'=\frac{2c}{g}\;
\frac{2\,sin\phi}{(\cos\phi)^2}
\]
which then determines the elapsed proper time for the ground clock at reunion with the light ray 
\begin{equation}
\tau_C=\frac{c}{g}\;\sinh^{-1}\left(\frac{2 \sin\phi}{\cos^2\phi}\right)
\end{equation}
This proper time is simply obtainable from equation (16) by taking the limit of a free body which travels at the speed of light. The second observable is light ray launch angle which results in a horizontal location for the reunion of light ray and the ground equal to that for the rod
\[
\tan\phi=\frac{gL}{2c^2}
\]
There is, of course, no elapsed proper time for a `clock on a light ray'.

\begin{figure}
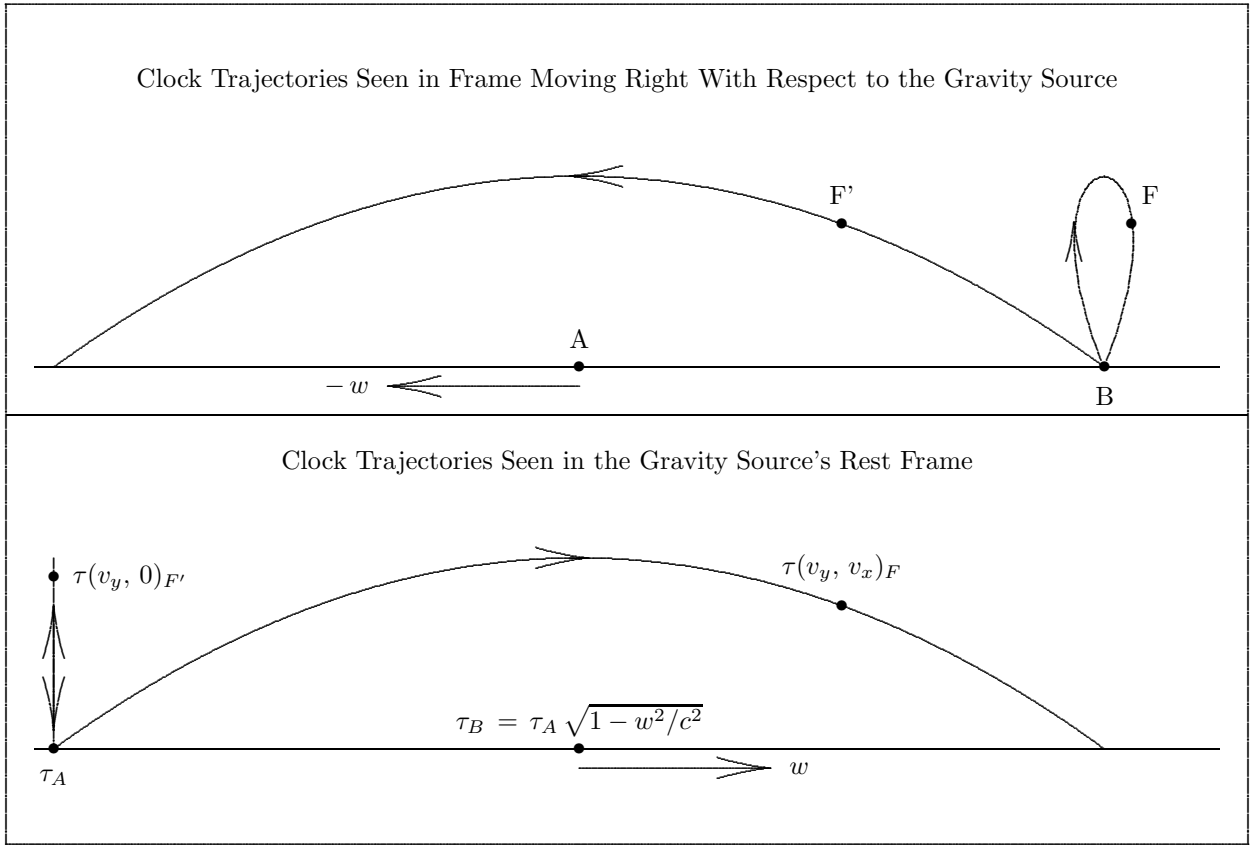

\beginpicture
\setcoordinatesystem units <1in,1in> point at 0 0
\setplotarea x from -3.25 to 3.25, y from -1 to 3.91
\plotheading {\Large \bf Clock Trajectories Seen in Two Frames of Reference}
\setquadratic \plot -3 0 -.25 1 2.5 0 /
\setquadratic \plot -3 2 -.25 3 2.5 2 /
\put {$\bullet$} at -.25 0
\put {$\tau_B\,=\,\tau_A\,\sqrt{1-w^2/c^2}$} at -.25 .15
\put {$\bullet$} at -.25 2
\put {A} at -.25 2.15
\put {$\bullet$} at -3 0
\put {$\tau_A$} at -3 -.15
\put {$\bullet$} at 2.5 2
\put {B} at 2.5 1.85
\put {$\bullet$} at 2.64 2.75
\put {F} at 2.74 2.9
\put {$\bullet$} at 1.125 .75
\put {$\tau(v_y,\,v_x)_F$} at 1.125 .95
\put {$\bullet$} at 1.125 2.75
\put {F'} at 1.125 2.9
\put {$\bullet$} at -3 .9
\put {$\tau(v_y,\,0)_{F'}$} [l] at -2.9 .9
\setlinear \plot -3 .01 -3 1 /
\putrule from -3.1 0 to 3.1 0
\putrule from -3.1 2 to 3.1 2
\putrule from -3.25 1.75 to 3.25 1.75
\setlinear \plot -3.25 -.5 3.25 -.5 3.25 3.9 -3.25 3.9 -3.25 -.5 /
\put {Clock Trajectories Seen in the Gravity Source's Rest Frame} at 0 1.5
\put {Clock Trajectories Seen in Frame Moving Right With Respect to the Gravity Source} at 0 3.5
\arrow <20pt> [.15,.4] from -.25 -.1 to .75 -.1
\arrow <20pt> [.15,.4] from -.25 1.9 to -1.25 1.9
\arrow <20pt> [.15,.4] from -.3 1 to -.2 1
\arrow <20pt> [.15,.4] from -.2 3 to -.3 3
\arrow <20pt> [.15,.4] from -3 .5 to -3 .75
\arrow <20pt> [.15,.4] from -3 .5 to -3 .1
\arrow <15pt> [.15,.4] from 2.34 2.75 to 2.34 2.76
\put {$w$} [l] at .85 -.1
\put {$-\,w$} [r] at -1.35 1.9
\setcoordinatesystem units <.004in,.004in>
\setlinear \plot "geo.dat" 
\endpicture
\caption {Freely falling and ground clocks, marked $F,\;F',\;A,\;B$, in gravity are shown from two frames of reference --- the lower viewpoint is at rest with respect to the gravity source, and the upper viewpoint moves to the right at speed $w$ (source of gravity moves to left at speed $w$). In the frame in which gravity's source is at rest, clock $F'$ is launched into vertical free fall, and ground clock $A$ waits at rest for the reunion.  In the same frame clock $F$ is on a free falling trajectory which moves to the right, and clock $B$ moves on the ground at constant velocity to the right to meet the return of $F$ to the ground.  Proper times at the various reunions of these clocks and other related observables are calculated in the gravity-free, but accelerating {\it ground floor} situations; and fulfillment of the SREP requires that those results must all be reproduced in each of these two illustrated situations in gravity (the four clock times are shown in the lower view). This specifies modifications of the gravitational equations of motion when these equations are stated in the frame moving with respect to the source of the gravity, including {\it gravitomagnetic} terms which are proportional to the velocity of the gravity's source.}
\end{figure}

\subsection*{Moving Gravity Source}

Trajectories of these various clocks and the light ray in the gravity environment are shown in Figure 5. The lower picture gives the scene in the rest frame of the gravity source and the {\it at rest} ground clocks. The upper picture gives the scene in the frame of the clock which moves to the right so as to record the reunion of the freely falling clock launched to the right. When the unusual motion of the freely falling clock in the upper scene is required to also occur in gravity, additional  {\it gravitomagnetic-like} acceleration terms to the freely falling clock's equation of motion are required which are in proportion to the motion of the gravity source in that frame of observation.

Another consequence of performing measurements in the frame moving with the ground clocks at speed $w$ is a change in the measured local value of gravitational acceleration. Since this moving clock will experience an elapsed proper time smaller than that of the ground clocks at rest
\begin{equation}
\tau(w)_B\;=\;\tau_A\;\sqrt{1-w^2/c^2}
\end{equation}
and the vertical speed of the launched freely falling body is enhanced as measured in this frame (time dilation)
\[
v_y'\;=\;\frac{1}{\sqrt{1-w^2/c^2}}\;v_y
\]
observers accompanying the horizontally moving ground clocks record a local gravitational acceleration of
\[
g(w)\;=\;\frac{1}{1-w^2/c^2}\;g\;\cong\;\left(1+w^2/c^2\right)\;g
\]
The SREP's enforcement of this will require further modifications of the gravitational equation of motion when expressed in frames in which the source of gravity is moving.

A clock launched vertically in the frame which is not horizontally moving relative to gravity's source can also be viewed from the rest frame of the ground clocks which travel to the right. The elapsed proper time for the freely falling clock in that case is obtained from equation (15) with $v_x=0$
\[
\tau (v_y,0)_{F'}\;=\;\frac{2v_y}{g}\;\frac{1}{\sqrt{1-v_y^2/c^2}}
\]

In the frame of reference moving horizontally along the ground at speed $w$, this situation is seen as a clock launched into gravitational free fall and moving to the left, along with another clock also moving to the left along the ground such that it meets the freely falling clock at its reunion with the ground. A further {\it gravitomagnetic} acceleration term will be required to obtain equivalent observational outcomes when this situation is considered in gravity.

\section*{Requirements for Equivalent Predictions in Gravity}

All the phenomena and situations considered in the preceding sections must be considered again in an environment of real gravitational acceleration $g$ as measured on the ground. The outcomes for all the observables previously obtained by kinematical calculations in gravity-free space must be reproduced under identical arrangements in the gravity environment if the SREP is valid. To achieve this, $1/c^2$ order gravitational corrections to the equations of motion for freely falling bodies, to the expression for the {\it proper} tick rates of clocks in gravity, and to the speed of light function are required \cite{fn1}. Expressing each of these three equations in terms of a proper time variable $\tau$ which represents the elapsed time of clocks at rest, the modified rate for clocks in general motion and at general altitude above the ground is assumed to be
\begin{equation}
d\tau(\vec{r},\;\vec{v})=d\tau\;\left(1-\frac{1}{2}\frac{v^2}{c^2}+a_1\frac{\vec{g}\cdot\vec{r}}{c^2}\right)
\end{equation}
The equation of motion for bodies freely falling in the gravity is assumed to be
\begin{equation}
\frac{d^2\vec{r}}{d\tau^2}=\vec{g}\left(1+a_2\frac{v^2}{c^2}+a_3\frac{\vec{g}\cdot\vec{r}}{c^2}+a_5\frac{(\hat{g}\cdot\vec{v})^2}{c^2}\right)+a_4 \frac{\vec{g}\cdot\vec{v}\vec{v}}{c^2}
\end{equation}
with $\vec{v}=d\vec{r}/d\tau$. And the light speed function in gravity is assumed to be
\begin{equation}
c(\vec{r})\;=\;\frac{|d\vec{r}|}{d\tau}\;=\;c\:\left(1\:+\:a_6\frac{\vec{g}\cdot\vec{r}}{c^2}\right)
\end{equation} 
{\bf Values for the numerical coefficients in these three equations, $a_1 ...a_6$, are sought so that the observables previously obtained kinematically in gravity-free inertial space are reproduced in the corresponding situation in gravity.}  A unique solution will result.

A freely falling clock is launched with the same initial velocity used previously  --- $(v_x, v_y)$.  The horizontal equation of motion from equation (47) is first considered:
\[
\frac{d^2 x}{d\tau^2}=-a_4\; \frac{g}{c^2}\frac{dy}{d\tau}\frac{dx}{d\tau}
\]
Since the right hand side of this equation is already proportional to $1/c^2$, the Newtonian trajectory for the freely falling body
\[
x(\tau)=v_x \tau \hspace{.5in}y(\tau)=v_y \tau -\frac{1}{2} g \tau^2
\]
can be employed in its evaluation. Integrating this horizontal equation of motion, demanding that the distance given in equation (11) is reached at proper time given by equation (15):
\[
\frac{2v_x v_y}{g}\frac{1}{1-v_y^2/c^2}=v_x\tau_G+\int_0^{\tau_G} d\tau\int_0^{\tau}\frac{d^2 x}{d\tau^2}\:d\tau \;\mbox{ to order }\; 1/c^2
\]
requires
\[
a_4=-2
\]
The vertical equation of motion from equation (20) is now considered:
\begin{equation}
\frac{d^2 y}{d\tau^2}=-g\left(1+a_2\frac{(dy/d\tau)^2+(dx/d\tau)^2}{c^2}-a_3\frac{gy}{c^2}+(a_5-2)\frac{(dy/d\tau)^2}{c^2}\right)
\end{equation}
in which the result for $a_4$ has been incorporated. Since the proper time for the reunion of clock and ground as recorded by the ground clock is given by equation (16) and is independent of the horizontal speed of the body. This result can only emerge when solving equation (22) if
\[
a_2=0
\]
The remaining dimensionless coefficients in equation (20) are fixed by using the Newtonian motion on the right hand side, integrating from the initial vertical position $0$ and speed $v_y$, and requiring both the return of the freely falling clock to the ground and the reversal of the vertical velocity to $-v_y$ to occur at time $\tau_A$. This yields
\[
a_3=-1 \hspace{.5in} a_5=0
\]

The proper time of the reunion with the ground as recorded by the freely falling clock is obtained by integrating the clock rate expression given in equation (19). Demanding that the result be equal to the kinematically derived amount given in equation (15) yields the value of the 'red shift' coefficient in equation (19)
\[
a_1=-1
\]
It should be noted that this derivation of the gravitational 'red shift' of clock rates did not employ light ray propagation between differently located clocks.  Combining these results for the clock rate expression and the equation of motion expression, their coefficients now determined, the locally measured acceleration rate for a body instantaneously at rest is found to be dependent on altitude 
\[
g(y)_{\rm local}=g\left(1-\frac{gy}{c^2}\right)
\]

In the limit of small initial elevation angles, light rays move in gravity along the curves
\[
y(x)=\frac{g}{2c^2}\;x(x_{\gamma}-x)
\]
The proper elapsed time of ground clocks for the reunion of the light ray with the ground has already been determined  and is given in equation (17). Demanding this same elapsed proper time in gravity, the light ray speed function is assumed , and integration over the light trajectory is performed to obtain total elapsed time. Corrections to the light trajectory of order $1/c^2$ need not be considered as they will generate only $1/c^4$ order corrections to the result. Therefore 
\[
\tau_C=\int_{(0,0)}^{(x_{\gamma},0)}\frac{\sqrt{dx^2+dy^2}}{c(\vec{r})}
\]
This is fulfilled for the coefficient value
\[
a_6=-1
\]
which appears in the light speed function, equation (21). Combining this result with the clock rate expression given by equation (19), the locally measured speed of light is found to be independent of altitude in gravity.

In conclusion: {\bf In a frame of reference at rest with respect to a source of gravity which locally (at the ground) produces a gravitational acceleration $g$ and speed of light $c$ as measured by clocks at rest on the ground, then equivalence of all local phenomena to that which occurs in an accelerated but force-free environment requires the following $1/c^2$ order modifications to the local equations of motion for bodies, clocks, and light} \cite{fn5}
\begin{eqnarray}
\frac{d^2\vec{r}}{d\tau^2}\;&=&\;\vec{g}\;\left(1-\frac{\vec{g}\cdot\vec{r}}{c^2}\right)
\;-\;2\:\frac{\vec{g}\cdot\vec{v}\vec{v}}{c^2}  \\
d\tau(\vec{r},\:\vec{v})\;&=&\;d\tau\;\left(1-\frac{1}{2}\frac{v^2}{c^2}-\frac{\vec{g}\cdot\vec{r}}{c^2}\right)
 \\ c(\vec{r})\;&=&\;c\;\left(1-\frac{\vec{g}\cdot\vec{r}}{c^2}\right)
\end{eqnarray}

\subsection*{Geometrical Interpretation}

This body equation of motion is obtainable from the particle lagrangian
\[
L\;=\;\frac{1}{2}\:v^2\:\left(1\;+\;\frac{1}{4}\frac{v^2}{c^2}\right)\;+\;\vec{g}\cdot\vec{r}\:\left(1\;+\;\frac{1}{2}\frac{v^2}{c^2}\right)
\]
which to the exhibited $1/c^2$ order is equivalent to a {\it geometrical} least action principle
\begin{equation}
\delta A\;=\;0\;=\;\delta\:\int\sqrt{g_{\mu\nu}dx^{\mu}dx^{\nu}}\hspace{.5in}\mu,\:\nu=t,x,y,z
\end{equation}
with the dominant {\it time-time} component of the metric tensor being altered from the  Minkowski metric value $\eta_{tt}=1$, while the spatial values remain unchanged; $\eta_{xx}=\eta_{yy}=\eta_{zz}=-1/c^2$
\[
g_{tt}\;=\;(1-\vec{g}\cdot\vec{r}/c^2)^2
\]
The light speed function given by equation (25) then follows from the {\it null-geodesic} assumption
\begin{equation}
g_{\mu\nu}dx^{\mu}dx^{\nu}\;=\;0\hspace{.5in}for\;light
\end{equation}
and the clock rate equation (24) is the lagrangian invariant
\[
d\tau\;=\;\sqrt{g_{\mu\nu}dx^{\mu}dx^{\nu}}
\]

\subsection*{Moving Gravity Source}

An equation of motion for freely falling bodies which is valid in more general frames in which the source of the gravity moves would be informative. In this situation additional acceleration terms must be considered which are functions of the gravity source's velocity $\vec{v}_s$. By considering the previous phenomena from a reference frame which moves at constant velocity along the ground, three such terms can be determined
\begin{equation}
\delta\left(\frac{d^2\vec{r}}{d\tau^2}\right)\;=\;
\frac{1}{c^2}\;\left(\;a_7\;\vec{g}\cdot\vec{v}\,\vec{v}_s\;+\;a_8\;
\vec{v}\cdot\vec{v}_s\;\vec{g}\;+\;a_9\;v_s^2\:\vec{g}\;\right)
\end{equation}
Because the source velocity $\vec{v}_s$ is orthogonal to the local gravity direction in these situations, a number of other possible acceleration terms proportional to $\vec{g}\cdot\vec{v}_s$ are not brought into play and so remain undetermined by these SREP arguments.  

In the case of the clock originally launched up and to the right, with a ground clock following along the ground so as to arrive at the reunion of clock with ground, we recall that in the frame which follows the ground clock, the freely falling clock was launched not vertically upward, but with angle to the vertical of $\Theta\cong -2v_xv_y/3c^2$ (to leading order in $1/c^2$). As illustrated in Figure 5, it then moved on a closed trajectory which finished at its starting point on the ground. Demanding this outcome from the x-component of our body's equation of motion  with source of gravity moving to the left, then requires  
\[
-\frac{v_xv_y}{c^2}\frac{2v_y}{g}
\;+\;a_7\:\int_0^{2v_y/g}dt\int_0^{2v_y/g}\frac{wg}{c^2}
\left(v_y-gt\right)\:dt\;\cong\;0
\]
with $w\cong v_x$. This requires $a_7=2$. And as previously indicated, the vertical acceleration in this frame is not $g$, it is  $g(w)=g\left(1+w^2/c^2\right)$ which requires $a_9=1$.

If the case of the clock vertically launched in the original frame is now considered in the frame moving to the right at speed $w$, the vertical speed with which it was launched is $v_y(1-w^2/2c^2)$ to lowest order in $1/c^2$, while the total proper time for the ground clock traveling to the left to meet the freely falling clock at reunion with the ground is as given in equation (18).  Since this moving clock's proper time runs at a rate slower than that of the ground clocks at rest in this frame by the factor 
\[
\frac{d\tau}{d\tau(w)}\;=\;\sqrt{1-w^2/c^2}\;\cong\;1-\frac{1}{2}
\frac{w^2}{c^2}
\]
the vertical acceleration of the freely falling clock must be $g\,(1-w^2/c^2)$
to lowest order in $w^2$.  This fixes the final coefficient in equation (28) to be $a_8=-2$.     

The entire equation of motion, equation (28) plus the contributions from equation (23), is then
\begin{equation}
\frac{d^2\vec{r}}{d\tau^2}\;=\;\vec{g}\;\left(1\;-\;
\frac{\vec{g}\cdot\vec{r}}{c^2}\;+\;\frac{v_s^2}{c^2}\right)\;-\;\frac{2}{c^2}\,
\vec{g}\cdot\vec{v}\:\vec{v}\;+\;\frac{2}{c^2}\,\vec{v}\times\left(\vec{v}_s\times\vec{g}\right)
\end{equation}

A moving gravity source also changes the speed of light function. A Lorentz transformation to the frame traveling to the right at speed $w$ relates the launch angle of the light ray which will be seen in this frame to the original launch angle
\begin{eqnarray}
\tan\phi'\;&=&\;\frac{\sin\phi\:
\sqrt{1-w^2/c^2}}{\cos\phi\:-\:w/c}\hspace{.4in}or\;for\;small\;angles
\nonumber \\ \phi'\;&\cong&\;\phi\:\left(1+w/c\right) \nonumber
\end{eqnarray}
But the maximum height above the ground which the light ray reaches is unchanged by this transformation and is given approximately by 
\[
h\;\cong\;\frac{1}{2}\:\phi^2\;\left(\frac{dc}{cdy}\right)^{-1}
\]
The gravitomagnetically modified light speed function
\begin{equation}
c(\vec{r})\;=\;c\:\left(1\;-\;\frac{\vec{g}\cdot\vec{r}}{c^2}\left(1-2\hat{c}\cdot
\vec{v}_s/c\right)\right)
\end{equation}
is required to achieve this equivalent result; $\hat{c}$ is the unit vector in the direction of light propagation and again $\vec{v}_s$ is the velocity of the source of gravity.

These SREP results for a moving source are in agreement with what one obtains by applying a Lorentz transformation to the metric field previously found in the gravity source's rest frame.  From the transformation rule for a second rank tensor
\[
g_{\mu\nu}'\;=\;\sum_{\alpha}\sum_{\beta}
\frac{\partial x^{\alpha}}{\partial x'^{\mu}}
\frac{\partial x^{\beta}}{\partial x'^{\nu}}\;g_{\alpha\beta}
\]      
and the lowest order expression of the Lorentz transformation
\begin{eqnarray}
\vec{r}\;&\cong&\;\vec{r}\,'\;-\;\vec{v}_s t' \nonumber \\
t\;&\cong&\;t'\;-\;\vec{v}_s\cdot\vec{r}\,'/c^2 \nonumber
\end{eqnarray}
a spatial vector of (mixed time-space) metric components is obtained
\[
g_{io}'\;=\;g_{oi}'\;\cong\;2\:\vec{g}\cdot\vec{r}\;\left(\vec{v}_s\right)_i/c^4\hspace{.5in}components\;i=x,y,z
\] 
which inserted into the action principle given by equation (26) generates the new lagrangian term
\begin{equation}
\delta L\;=\;-\:2\:\vec{g}\cdot\vec{r}\:\vec{v}_s\cdot\vec{v}/c^2
\end{equation}

\section{Periastron Precession}

Just about any modification from an inverse square central acceleration law causes the major axis of Keplerian  orbits to precess in inertial space. In particular.  This holds, in particular, for the modifications to the equation of motion which result from the SREP as given by equation (23). Consider a body which is close to being in circular orbit around a central body. Small perturbations are considered from the mean circular motion so that the time evolution of the {\it eccentric} deviations from circularity can be derived and compared to the mean orbital motion. Starting with the radial and tangential equations of motion
\begin{eqnarray}
\frac{d^2r}{d\tau^2}\;&=&\;\vec{g}(\vec{r},\vec{v})\cdot\hat{r}\;+\;\omega^2\:r \nonumber\\
\frac{d}{d\tau}\left(r^2\omega\right)\;&=&\;r\:\vec{g}(\vec{r},\vec{v})\cdot\hat{t} \nonumber
\end{eqnarray}
small perturbations are considered about a circular orbit, $r\rightarrow r+x(\tau),\;\omega\rightarrow \omega+\delta\omega(\tau)$.  The needed acceleration components from equation (23) are 
\begin{eqnarray}
\vec{g}\cdot\hat{r}\;&=&\;-g\:\left(1+gx/c^2\right) \nonumber \\
\vec{g}\cdot\hat{t}\;&=&\;\frac{2gv}{c^2}\:\frac{dx}{d\tau} \nonumber
\end{eqnarray}
with $v$ being the horizontal velocity of the mean circular orbit. The linearized equation for radial perturbation $x(\tau)$ then becomes
\[
\frac{d^2x}{d\tau^2}\;+\;\left(3\omega^2+dg/dr-3g^2/c^2\right)\;x\;=\;0
\]
with the radial tidal gradient of the solar system's total acceleration field $dg/dr$ added, and the relationship $g=v\omega$ being used to simplify the radial and tangential $1/c^2$ order perturbation terms.  The resulting radial perturbation is simple {\it eccentric} harmonic motion with arbitrary amplitude and phase determined by initial conditions.  But this eccentric motion's frequency $\omega_o$ is specific, and relative to the orbital frequency it is shifted by the SREP modifications of the dynamics to be slightly less than what it will be due solely to the tidal gradient $dg/dr$ This increased frequency difference between orbital and eccentric motions appears in space as an addition to the total precession rate of the orbit's major axis in the positive sense of the orbital motion (prograde precession), and of amount
\[
\delta\left(\omega-\omega_o\right)\;\cong\;\frac{3}{2}\:g^2/(\omega c^2)\;\cong\;\frac{3}{2}\frac{v^2}{c^2}\;\omega
\]

Prior to Einstein's development of his special relativity theory in 1905 and formulation of his Equivalence Principle beginning in 1907, a century of astronomical observations had already discovered about a $43\; arc-seconds/century$  precession rate of Mercury's orbit in excess of what could be understood from consideration of the Newtonian perturbations by the other known planets of the solar system. Half this anomalous precession  is here accounted for from the SREP 
\begin{equation}
\left(\frac{3}{2}\,\frac{v^2}{c^2}\;\omega\right)_{Mercury}\;\cong\;22.5\;arc-sec/century
\end{equation}  

\subsection*{A Historical Speculation}

As early as December 1907 Einstein mentioned in a letter to a friend that, "I am now occupied with a relativistic treatment of the law of gravity, with which I hope to explain the anomalous secular change in the perihelion of Mercury." And he added in a footnote, "Up to now the thing doesn't appear to want to succeed." \cite{fn2} Had Einstein arrived at the SREP's prediction, equation (32), about this time?  By then he certainly was in a position to extend his EP to a full SREP. Perhaps he had done so but chose to not publish the consequences of a full special relativistic generalization of his principle because this perihelion prediction was only half the {\it known} anomaly in Mercury's orbital motion?  Yes, his prediction of light deflection from the EP was also only half that which would eventually emerge from his complete gravity theory of 1915/16, but in 1907 neither the full theory's prediction for light deflection nor its experimental measurement during the eclipse of 1919 were available to create a conflict. 

On the other hand, continued work toward a complete relativistic theory of gravity may have been spurred on by such an anomalous early EP-inspired estimate which produced contributions to Mercury's perihelion precession rate with magnitude being a simple fraction of the observed anomaly of $43\;arc-sec/century$.  From several letters from Einstein to colleagues written around the end of 1915, Einstein mentioned that one of the things which had kept him searching right up until the end for a better metric tensor theory of gravity was that his "old theory" only explained half Mercury's anomalous perihelion precession.  And then when he recalculated this effect in November 1915 using the new vacuum field equations of his final metric tensor theory of general relativity,  and did obtain the full anomaly ``without any special hypothesis'', he mentioned in another letter that this produced one of the strongest emotional experiences of his career, ``... for a few days I was beside myself with joyous excitement.''  It appears clear that the Mercury orbit anomaly played a continuous and key role in Einstein's search for a new theory of gravity.  Many narratives of this scientific revolution seem to have minimized this part of the story, and the focus on the later confirmation of the theory with the measurement of the deflection of light during the 1919 eclipse further overshadowed the perihelion precession phenomenon. 

\section{Summary}

Incorporating the special relativity theory more fully into Einstein's principle of equivalence between the phenomena in accelerated frames of reference and that in local gravitational fields has led to prediction of a number of additional effects in post-Newtonian gravity.  These include {\it geodetic precession} of local inertial frames which follow non-radial, free falling trajectories through gravity, precession of Mercury's perihelion, and {\it gravitomagnetic} forces between matter proportional to the velocities of both source matter and acted-upon matter, as well as {\it gravitomagnetic} precession. And the original predictions of Einstein's EP are, of course, also predicted --- universal reduction of clock rates  and both deflection and slowing of light in gravity.  

The SREP predictions generally do not account for the entire physical effects which are now routinely measured by experiments. Within the general class of locally Lorentz-invariant, complete metric theories of gravity --- all of which fulfill the SREP --- a variety of calculated post-Newtonian gravitational effects are tabulated below and expressed in terms of two dimensionless parameters, $\gamma$ and $\beta^*=2\beta-1$, which identify and quantify post-Newtonian features of the metric theories which go beyond the local physics specified by the equivalence principles.
\begin{eqnarray}
d\tau\;&=&\;dt\;
\left(1-\frac{1}{2}\frac{v^2}{c^2}-{\bf 1}\frac{\vec{g}\cdot\vec{r}}{c^2}\right) \nonumber\\
c(\vec{r})\;&=&\;c\;
\left(1-({\bf 1}+\gamma)\frac{\vec{g}\cdot\vec{r}}{c^2}\right) \nonumber \\
\vec{\Omega}_{geo}\;&=&\;\left({\bf 1/2}+\gamma\right)\;\frac{\vec{g}\times\vec{v}}{c^2} \nonumber \\
\Omega_{Merc}\;&=&\;\left({\bf 3/2}+2\gamma-\beta^*/2\right)\;\frac{v^2}{c^2}\:\omega  \nonumber \\
\vec{a}_{grav-mag}\;&=&\;
({\bf 2}+2\gamma)\;\frac{\vec{v}\times(\vec{g}\times\vec{v}_s)}{c^2} \nonumber \\
\frac{M_G}{M_I}\;&=&\;1\,+\,({\bf 1}+\gamma-2\beta^*)\frac{1}{2Mc^2}\int\frac{\rho(\vec{x})\rho(\vec{y})}{|\vec{x}-\vec{y}|}\,d^3x\,d^3y \nonumber
\end{eqnarray}
SREP contributions are shown in bold numbers.  These parameterized post-Newtonian (PPN) expressions for different (albeit theoretically connected) gravitational effects have been known for decades \cite{n68,n72,cwb}; indeed, it was my awareness of the contributions to these several phenomena which were independent of the specifics of the particular metric theory that motivated this investigation.  It is the SREP which dictates these universal contributions to post-Newtonian gravity.

{\bf This work was performed with support by National Aeronautics and Space Administration contract NASW-00011 and grant NAG8-1811.}

\subsection*{Appendix C:  Relationship to Complete Theories}

Beginning with an underlying metric field theory of gravity which is locally Lorentz-invariant, a $1/c^2$ order, N-body lagrangian can generally be derived.  The part of this lagrangian which is independent of the specifics of the metric theory and which manifests both local Lorentz invariance and the EP is
\begin{eqnarray}
L_{SREP}\;&=&\;\sum_i\left(\frac{1}{2}\,m_i\,v_i^2\;+\;\frac{1}{8c^2}\,m_i\,v_i^4\right) 
\;+\;\frac{G}{2}\;\sum_{i,\,j}\frac{m_i\,m_j}{r_{ij}}\left(1\,-\,\frac{1}{2c^2}\left(\vec{v}_i\cdot\vec{v}_j+\vec{v}_i\cdot\hat{r}_{ij}\hat{r}_{ij}\cdot\vec{v}_j\right)\right) \nonumber \nonumber \\
&+&\;\frac{G}{4c^2}\;\sum_{i,\,j}\frac{m_i\,m_j}{r_{ij}}\left(\vec{v}_i-\vec{v}_j\right)^2 \nonumber
\end{eqnarray}
with the first line by itself being Lorentz-invariant to $1/c^2$ order, but the additional Lorentz-invariant term on the second line being also needed in order to fulfill the EP.  Focusing on one of the N bodies in the presence of N-1 other quasi-static sources of  gravity seen by the selected body, one can expand this lagrangian about a chosen origin, rescale the time variable into the proper time variable at this origin, and then reproduce the SREP-derived equation of motion given by equation (61).  Giving  the source bodies motions $\vec{v}_s$, the SREP-derived gravitomagnetic equation of motion corrections found in equation (71) can also be obtained.

But there is additional $1/c^2$ order gravitational physics beyond the SREP.  It results from two lagrangian terms
\begin{eqnarray}
L_{\gamma}\;&=&\;\gamma\;\frac{G}{2c^2}\;\sum_{i,\,j}\frac{m_i\,m_j}{r_{ij}}\left(\vec{v}_i-\vec{v}_j\right)^2 \nonumber \\
L_{\beta^*}\;&=&\;-\,\beta^*\;\frac{G^2}{2c^2}\;\sum_{i,\,j,\,k}\frac{m_i\,m_j\,m_k}{r_{ij}\,r_{ik}} \nonumber
\end{eqnarray}
with the indices $i,\;j,\;k$ each being summed over the $N$ bodies \cite{n85}.  The two new coupling strength parameters have special values in General Relativity, $\gamma_{GR}=1$ and $\beta^*_{GR}=1$ ($\beta=(1+\beta^*)/2$ is the more traditional PPN, {\it Eddington} coefficient) \cite{e20}, but they individually have different values in scalar-tensor metric theories, for example.
In addition to contributing to additional gravitomagnetic interaction, the lagrangian term $L_{\gamma}$ produces a global non-Euclidean geometry for the arena of physical events and objects.  But locally this deviation from the Euclidean nature of space can be {\it delayed}. At a chosen locality $\vec{r}_o$, a sequence of spatial coordinate transformations involving first a rescaling of the spatial coordinates
\[
\vec{x}\,'\;=\;\left(1\,+\,\gamma\,U(\vec{r}_o)/c^2\right)\,\vec{x}
\]
with $U(\vec{r}_o)$ being the Newtonian potential at $\vec{r}_o$ of the gravitational sources, and $\vec{x}\,=\,\vec{r}-\vec{r}_o$; and then the non-linear {\it warping} of the coordinates 
\[
\vec{x}\,'\;=\;\vec{\rho}\;+\;\frac{1}{2c^2}\gamma \vec{g}\rho^2\;-\;\frac{1}{c^2}\gamma
\vec{g}\cdot\vec{\rho}\vec{\rho}
\]
the locality only experiences the onset of non-Euclidean spatial effects at the quadratic order in laboratory size.  The non-linear lagrangian term $L_{\beta*}$ produces three-body gravitational interactions, and it also produces modifications to the gravitational potential between two bodies whose strength is proportional to the square of one mass or the other and depends on the inverse square of body separation.  Neither of these two lagrangian terms can be inferred by SREP arguments; a full field theory of gravity is required for their specification.

\end{document}

%% file: prepicte.tex
\catcode`@=11 \catcode`!=11

\expandafter\ifx\csname fiverm\endcsname\relax
  \let\fiverm\fivrm
\fi
  
\let\!latexendpicture=\endpicture 
\let\!latexframe=\frame
\let\!latexlinethickness=\linethickness
\let\!latexmultiput=\multiput
\let\!latexput=\put
 
\def\@picture(#1,#2)(#3,#4){%
  \@picht #2\unitlength
  \setbox\@picbox\hbox to #1\unitlength\bgroup 
  \let\endpicture=\!latexendpicture
  \let\frame=\!latexframe
  \let\linethickness=\!latexlinethickness
  \let\multiput=\!latexmultiput
  \let\put=\!latexput
  \hskip -#3\unitlength \lower #4\unitlength \hbox\bgroup}

\catcode`@=12 \catcode`!=12

%% file: postpict.tex
\catcode`@=11 \catcode`!=11
  
\let\!pictexendpicture=\endpicture 
\let\!pictexframe=\frame
\let\!pictexlinethickness=\linethickness
\let\!pictexmultiput=\multiput
\let\!pictexput=\put

\def\beginpicture{%
  \setbox\!picbox=\hbox\bgroup%
  \let\endpicture=\!pictexendpicture
  \let\frame=\!pictexframe
  \let\linethickness=\!pictexlinethickness
  \let\multiput=\!pictexmultiput
  \let\put=\!pictexput
  \let\input=\@@input   
  \!xleft=\maxdimen  
  \!xright=-\maxdimen
  \!ybot=\maxdimen
  \!ytop=-\maxdimen}

\let\frame=\!latexframe

\let\pictexframe=\!pictexframe

\let\linethickness=\!latexlinethickness
\let\pictexlinethickness=\!pictexlinethickness

\let\\=\@normalcr
\catcode`@=12 \catcode`!=12

%% file: Dragone.bbl
\begin{thebibliography}{99}
\bibitem{fn1} Because the rod is moving essentially straight up or straight down in the instantaneous rest frames of the observer at each end of the rod's free flight, a horizontal rod experiences no reorientations due to Lorentz contractions; this simplifies the analysis of the gravitomagnetic precession.
\bibitem{fn4} $\sqrt{X^2+Y^2}$ is not the {\it proper} length of the rod; the coordinates $X,Y$ record the Lorentz-contracted position of the rod's end "2" at the  instant that end "1" has coordinates $0,0$, and as seen from a frame in which the rod moves with velocity components $(v_x,\;v_y)$.
\bibitem{fn1} Several derived results for observables calculated in the accelerating, gravity free laboratory have been done exactly within special relativity (to all orders in $v/c$ and $gt/c$). They could therefore be ingredients for SREP predictions in ultra-relativistic contexts.  In this paper, however, these expressions are approximated to only the necessary order for being equated to the corresponding expressions for the same observables calculated to $1/c^2$ order in gravity.
\end{thebibliography}
